\documentclass[conference, onecolumn]{IEEEtran}
\IEEEoverridecommandlockouts
\usepackage{cite}
\usepackage{amsmath,amssymb,amsfonts,mathrsfs,mathtools}
\usepackage{graphicx}
\usepackage{textcomp}
\usepackage{xcolor}
\usepackage{algorithm}
\usepackage{algpseudocode}
\usepackage{caption}
\usepackage{subcaption}
\usepackage{theorem}
\usepackage{algorithmicx}
\usepackage{algpseudocode}

\theoremstyle{definition}

\usepackage{multirow}
\usepackage{listings}
\usepackage{mathtools}
\usepackage{hyperref}
\usepackage[dvipsnames]{xcolor}                              

\usepackage{tabularx}
\newcolumntype{L}{>{\centering \arraybackslash}X}

\definecolor{keywordcolor}{rgb}{0.0,0.0,0.6}
\definecolor{commentcolor}{rgb}{0.0,0.5,0.0}
\definecolor{stringcolor}{rgb}{0.58,0,0.82}

\def\BibTeX{{\rm B\kern-.05em{\sc i\kern-.025em b}\kern-.08em
    T\kern-.1667em\lower.7ex\hbox{E}\kern-.125emX}}

\newcommand{\un}[1]{\underline{#1}}

\newtheorem{defn}{Definition}
\newtheorem{thm}{{\cal T}heorem}
\newtheorem{cor}{Corollary}
\newtheorem{prop}{Proposition}
\newtheorem{lem}{Lemma}
\newtheorem{conj}{Conjecture}
\newtheorem{constr}{Construction}
\newtheorem{note}{Remark}
\newtheorem{example}{Example}
\newcommand{\bit}{\begin{itemize}}
	\newcommand{\eit}{\end{itemize}}
\newcommand{\bcor}{\begin{cor}}
	\newcommand{\ecor}{\end{cor}}
\newcommand{\beq}{\begin{equation}}
	\newcommand{\eeq}{\end{equation}}
\newcommand{\beqn}{\begin{equation}}
	\newcommand{\eeqn}{\end{equation}}
\newcommand{\bea}{\begin{eqnarray}}
	\newcommand{\eea}{\end{eqnarray}}
\newcommand{\bean}{\begin{eqnarray*}}
	\newcommand{\eean}{\end{eqnarray*}}
\newcommand{\ben}{\begin{enumerate}}
	\newcommand{\een}{\end{enumerate}}
\newcommand{\bdefn}{\begin{defn}}
	\newcommand{\edefn}{\end{defn}}
\newcommand{\bnote}{\begin{note}}
	\newcommand{\enote}{\end{note}}
\newcommand{\bprop}{\begin{prop}}
	\newcommand{\eprop}{\end{prop}}
\newcommand{\blem}{\begin{lem}}
	\newcommand{\elem}{\end{lem}}
\newcommand{\bthm}{\begin{thm}}
	\newcommand{\ethm}{\end{thm}}
\newcommand{\bconj}{\begin{conj}}
	\newcommand{\econj}{\end{conj}}
\newcommand{\bconstr}{\begin{constr}}
	\newcommand{\econstr}{\end{constr}}
\newcommand{\bpf}{\begin{proof}}
	\newcommand{\epf}{\end{proof}}
\newcommand{\bprf}{{\em Proof: }}
\newcommand{\eprf}{\hfill $\Box$}

\newcommand{\und}[1]{\underline{#1}}
\newcommand{\lambdai}{\lambda^I}
\newcommand{\lambdaf}{\lambda^F}

\newcommand{\Null}{\mathcal{N}}

\begin{document}

\title{Constructions of Secure Convertible Codes}

\author{
  \IEEEauthorblockN{Manasa Suresh B P}
  \IEEEauthorblockA{IIT Hyderabad, India  \\ manasasureshbp22@gmail.com}
  \and 
  
  \IEEEauthorblockN{Vinayak Ramkumar}
  \IEEEauthorblockA{TU Munich, Germany \\
 vinram93@gmail.com}
  \and 
   \IEEEauthorblockN{Myna Vajha}
  \IEEEauthorblockA{IIT Hyderabad, India  \\ mynaramana@gmail.com}
  }

\maketitle
\begin{abstract}
    Secure convertible codes, introduced by Zhang and Rashmi, achieve secrecy capacity under a passive eavesdropper model in which an eavesdropper is allowed to access any \(\ell\) symbols from the initial and final codewords. In this paper, we construct two families of access-optimal secure convertible codes. The first family of codes based on Maximum Rank Distance (MRD) codes is constructed for all parameters in the merge and split regimes and requires a large field size. The second family of codes, constructed using Generalized Reed-Solomon codes, uses smaller fields but is applicable only for certain parameters in the merge and split regimes. 
\end{abstract}

\section{Introduction}\label{intro}


Maturana and Rashmi introduced convertible codes in \cite{MaturanaRashmi_ITCS2020} motivated by systems studies in  \cite{Kadekodi_osdi20, Kadekodi_Rashmi_GregoryGanger_clustersystems}, to enable efficient transformation of data from an $[n^I,k^I]$ initial code to an $[n^F,k^F]$ final code while minimizing the conversion cost.  
To enable the conversion, a message vector $\un{m}\in\mathbb{F}_q^{M}$ of size 
$M=\operatorname{lcm}(k^I,k^F)$ is considered.  The message is partitioned into $\lambdai=\frac{M}{k^I}$, $\lambdaf=\frac{M}{k^F}$  initial and final message blocks, respectively. Thus, the initial configuration consists of $\lambdai$ codewords, while the final configuration consists of $\lambdaf$ codewords. The message symbols encoded into each codeword are determined by the initial and final partitions 
$\mathcal{P}^I=\{\mathcal{U}_1,\mathcal{U}_2,\ldots,\mathcal{U}_{\lambdai}\}$ and $\mathcal{P}^F=\{\mathcal{V}_1,\mathcal{V}_2,\ldots,\mathcal{V}_{\lambdaf}\}$ of the message index set $[M]=\{1,\ldots,M\}$. The blocks within each partition are pairwise disjoint, and $\cup_{i=1}^{\lambdai} \mathcal{U}_i = \cup_{j=1}^{\lambdaf} \mathcal{V}_j = [M]$. $\mathcal{U}_i$ indicates the set of indices of the message symbols that are part of the $i$-th initial codeword and $\mathcal{V}_j$ indicates the set of indices of the message symbols that are part of the $j$-th final codeword. Therefore $|\mathcal{U}_i|=k^I$, $|\mathcal{V}_j| = k^F$.

An $[n^I,k^I;n^F,k^F]$ convertible  code over $\mathbb{F}_q$ consists of 
\ben
\item a pair of codes ($\mathcal{C}^I,\mathcal{C}^F$), where $\mathcal{C}^I$ is an $[n^I ,k^I]$ code and $\mathcal{C}^F$ is an $[n^F, k^F]$ code both defined over $\mathbb{F}_q$,
\item a pair of partitions $(\mathcal{P}^I, \mathcal{P}^F)$ of the message index set $[M]$, where every block $\mathcal{U}_i\in\mathcal{P}^I$  has size $k^I$ and every block $\mathcal{V}_j\in\mathcal{P}^F$ has size $k^F$, 
\item a conversion procedure  that takes $\lambdai$ initial codewords  from  ${\mathcal{C}^I}$ and converts them to $\lambdaf$ final codewords from ${\mathcal{C}^F}$.
\een

\begin{figure}[ht!]
    \centering
        \vspace{-0.15in}
\includegraphics[width=0.45\textwidth]{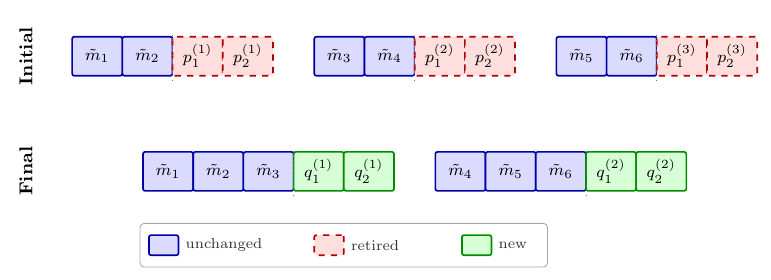}
    \caption{\small Illustration of $[n^I=4, k^I=2; n^F=5, k^F=3]$ convertible code where $\lambdai=3$ initial codewords are converted to $\lambdaf=2$ final codewords.} 
    \label{fig:convcode}
    \vspace{-0.1in}
\end{figure}

During the conversion procedure, the codeword symbols are classified into three types:  \emph{unchanged}, \emph{retired}, and \emph{new} symbols. An unchanged symbol is an initial codeword symbol that remains in the final codewords. A retired symbol is an initial code symbol that is removed during conversion, whereas a new symbol is generated during
conversion and added to the final configuration (see Fig.~\ref{fig:convcode}). A convertible code is said to be Maximum Distance Separable (MDS) if both the initial and final codes are MDS. The merge regime is a special class of convertible codes in which multiple initial codewords are converted into a single final codeword, i.e., $\lambdaf=1$, resulting in $k^F= \lambdai k^I$. The split regime corresponds to the case where a single initial codeword is transformed into multiple final codewords, and can be viewed as the dual of the merge regime. 

The efficiency of a convertible code is measured by the \emph{access cost} and \emph{bandwidth cost} of the conversion procedure. The access cost of a conversion procedure is the total number of code symbols accessed during the conversion procedure, and the bandwidth cost of the conversion procedure is the total amount of data read and written during the conversion procedure. 
\paragraph*{Prior Work}
In \cite{MaturanaRashmi_ITCS2020}, Maturana and Rashmi introduced the convertible code framework and constructed access-optimal MDS convertible codes in the merge regime. They introduced a matching lower bound on the access cost of MDS convertible codes in the merge regime. In \cite{MaturanaMukkaRashmi_ISIT20_LinearMDS, Maturana_Rashmi_TIT}, access-optimal MDS convertible code constructions are studied for all valid parameter settings along with matching access-cost lower bounds. In \cite{SaranshMaturanaRashmi, Convertible_ISIT25, Myna_TIT_MDS}, constructions of low-field-size MDS convertible codes in the merge regime are presented for all parameters. Bandwidth-optimal convertible codes and lower bounds are established in \cite{MaturanaRashmi_BWMerge_TIT, Tight_Lower_Bounds_MDS_Split_Singhvi_Chopra_Rashmi}. 
In \cite{Kong_LRC_TIT,MaturanaRashmi_LRC_ISIT23,Bandwidth_Cost_LRC_Rashmi}, convertible codes for locally repairable codes (LRCs) are studied with respect to the access and bandwidth cost. In \cite{MSRconvertible_isit26}, the authors study convertible minimum storage regenerating(MSR) codes.
In \cite{zhang_secure_25}, Zhang and Rashmi introduced an information-theoretic secrecy model for convertible codes in the presence of passive eavesdroppers and proposed constructions of access-optimal secure convertible codes that achieve the secrecy capacity. 

\paragraph*{Our Contributions} 
In this paper, we first make an observation that the secure convertible codes introduced in \cite{zhang_secure_25} are secure against an eavesdropper accessing only a subset of unchanged symbols, but not necessarily secure against an eavesdropper accessing a subset of unchanged and retired symbols. We propose two families of secure convertible codes that are secure against an eavesdropper accessing up to any $\ell$ of the unchanged, retired, and new symbols. Both these constructions are described through a precoding matrix $P$ and the generator matrices of initial and final codes of a base convertible code $\hat{G}^I, \hat{G}^F$. The first construction yields secure convertible codes for all valid parameters in the merge and split regimes. In this construction, $P$ is based on a generator matrix of a Maximum Rank Distance (MRD) code, and  $\hat{G}^I$ and $\hat{G}^F$ are allowed to be generator matrices corresponding to any access-optimal convertible code (see Theorem~\ref{thm:MRD} and Remark~\ref{rem:MRDdecod}). A drawback of this construction is that it requires a large field size. To circumvent this issue, for certain ranges of merge and split regimes, we construct a second family of secure convertible codes where $\hat{G}^I$ and $\hat{G}^F$ are chosen such that they are generator matrices of Generalized Reed-Solomon (GRS) codes (see Theorem~\ref{thm:GRSsecureconv}).  

\paragraph*{Notation}
Let $\mathbb{F}_q$ denote the finite field with $q$ elements. For integers $a\le b$, let $[a:b]=\{a,a+1,\ldots,b\}$ and for a positive integer $n$, let $[n]= [1:n]$. The $i$-th entry of a vector $\underline{x}$ is denoted by $x_i$. For an index set $\mathcal{B}\subseteq[n]$, the projection of $\underline{x}$ onto the coordinates indexed by $\mathcal{B}$ is denoted by $\underline{x}_{\mathcal{B}}=[x_i:i\in\mathcal{B}]$. For a matrix $G\in\mathbb{F}_q^{k\times n}$, the $(i,j)$-th entry is denoted by $G(i,j)$. The $i$-th row and $j$-th column of $G$ are denoted by $G(i,:)$ and $G(:,j)$, respectively. For row and column index sets $\mathcal{A}\subseteq[k]$ and $\mathcal{B}\subseteq[n]$, the corresponding sub-matrix is denoted by $G(\mathcal{A},\mathcal{B})$. Similarly, $G(:,\mathcal{B})$ and $G(\mathcal{A},:)$ denote the sub-matrices obtained by selecting the columns indexed by $\mathcal{B}$ and the rows indexed by $\mathcal{A}$, respectively. Capital letters (e.g., S, X) denote random variables, and lowercase letters (e.g., s, x) denote realizations.


\section{Preliminaries \label{sec:preliminaries}}

We now describe the framework of secure convertible codes introduced by Zhang and Rashmi in \cite{zhang_secure_25}. It extends the standard convertible code framework by incorporating an information-theoretic secrecy requirement against passive eavesdroppers while preserving the decoding and code-conversion properties. 
Specifically, an eavesdropper observing any $\ell$ codeword symbols across the initial and final configurations obtains no information about the secure message. 

Let $S$ denote a $k_S$-length random vector 
representing the secure message that is uniform over $\mathbb{F}_q^{k_S}$. 
We use $S_i^I$ and $S_j^F$ to denote the portions of the secure message associated with the $i$-th initial codeword and the $j$-th final codeword, respectively. Let $X^I, X^F$ be the random vectors representing the $n^I \lambdai$ initial and $n^F \lambda^F$ final codeword symbols respectively. Let $X_i^I$ and $X_j^F$ denote the $i$-th initial codeword and the $j$-th final codeword, respectively. For an index set $\mathcal{B}$ of codeword coordinates, $X_{i,\mathcal{B}}^I$ and $X_{j,\mathcal{B}}^F$ denote the projections of $X_i^I$ and $X_j^F$ onto the coordinates indexed by $\mathcal{B}$, respectively.  $X^I_{\mathcal{B}}$, $X^F_{\mathcal{C}}$ represent symbols of initial and final codewords respectively indexed by $\mathcal{B},\mathcal{C}$ respectively.

\bdefn[Secure Convertible Code \cite{zhang_secure_25}]\label{def:secconv}
An $(\ell, k_S)$-secure $[n^I, k^I; n^F, k^F]$ convertible code is an 
$[n^I, k^I; n^F, k^F]$ convertible code that can store a message $S$ that is uniform in 
$\mathbb{F}_q^{k_S}$ 
satisfying the following properties: 

\bit
\item Decoding property:  
For each $i \in [\lambdai]$, $j \in [\lambdaf]$ and any subset $\mathcal{B}_1 \subseteq [n^I]$ of size $k^I$, $\mathcal{B}_2 \subseteq [n^F]$ of size $k^F$,
\[
H\big(S_i^I \mid X_{i,\mathcal{B}_1}^I\big) = 0, \ H\big(S_j^F \mid X_{j,\mathcal{B}_2}^F\big) = 0.
\]
\item $\ell$-Secrecy:  
For any $\mathcal{E}^I \subseteq [\lambdai n^I]$, $\mathcal{E}^F \subseteq [\lambdaf n^F]$ such that 
$|\mathcal{E}^I| + |\mathcal{E}^F| \le \ell$: $
H\big(S \mid X_{\mathcal{E}^I}^I, X_{\mathcal{E}^F}^F\big) = H(S).$
\eit
\edefn

In \cite{zhang_secure_25}, the authors prove an upper bound on secrecy capacity, $k_S$ for any $(\ell, k_S)$-secure $[n^I, k^I; n^F, k^F]$ convertible code as: 
\bean
\label{eq:secretcapub}
k_S \le \min\{\lambdai (k^I - \ell), \lambdaf (k^F - \ell)\}.
\eean







\subsection{Necessary and Sufficient Conditions for Secrecy}



We now present a general setting that helps us describe the conditions required for the $\ell$-secrecy property of secure convertible codes in the next section. 
\bdefn An $[n, k]$ linear code is $(\ell, \hat{k})$-secure for $\hat{k} \le (k-\ell)$ if there exists a $k \times n$ generator matrix $G$ that encodes a secret message $\un{s} \in \mathbb{F}_q^{\hat{k}}$ using a key $\un{\kappa} \in  \mathbb{F}_q^{k - \hat{k}}$ as 
\bean
[\underline{s}~\underline{\kappa}] G = \un{c} \text{  where } G = [G_s^T~G_{\kappa}^T]^T,
\eean
$G_s$ is  a $(\hat{k} \times n)$ matrix 
and $G_{\kappa}$ is a $((k-\hat{k}) \times n )$ matrix such that an eavesdropper accessing $\ell$ codeword symbols of $\un{c}$ obtain no information about the secure message. Formally, it is required that $I(S; C_{\cal B})=0$ where $S, C$ are random vectors representing the secure message and codeword respectively and let $C_{\mathcal{B}}$ be the code symbols accessed by the eavesdropper for $\mathcal{B} \subseteq [n]$ such that $|\mathcal{B}|=\ell$.
\edefn
We now present the necessary and sufficient conditions required on a generator matrix for the code to be secure.



\blem \label{lem:secrecycondition}
An $[n,k]$ linear code with generator matrix $G= [G_s^T~G_{\kappa}^T]^T$ is $(\ell, \hat{k})$-secure iff rank($G_{\kappa}(:,\mathcal{B})) = j$ for any $\mathcal{B} \subseteq [n]$ such that $|\mathcal{B}| = j \le \ell$ and rank$(G(:,{\mathcal{B}}))=j$. 
\elem
\bprf
Let {$C = SG_s + K G_{\kappa}$}, where $S \in \mathbb{F}_q^{\hat{k}}$ is the random vector representing the secure message and $K \in \mathbb{F}_q^{k-\hat{k}}$ is a random vector representing the uniformly random key.
The observed symbols are  $ C_\mathcal{B} = SG_s(:,\mathcal{B}) + K G_{\kappa}(:,\mathcal{B}).$ The rank condition rank$(G(:,\mathcal{B}))=j$ ensures that the $j \le \ell$ symbols observed by eavesdropper are independent. Perfect secrecy follows iff probability, $Pr(S= \un{v} \mid C_\mathcal{B}= \un{u})=Pr(S=\un{v})$
for all $\un{v} \in \mathbb{F}_q^{\hat{k}}$. Using Bayes' rule,
\bea
\label{eq:bayes}
\scalebox{0.98}{$
Pr(S= \un{v} \mid C_\mathcal{B}= \un{u})
=
\frac{Pr(C_\mathcal{B}= \un{u} \mid S=\un{v})Pr(S=\un{v})}
{\sum_{\un{v}'} Pr(C_\mathcal{B}=\un{u} \mid S=\un{v}')Pr(S=\un{v}')}.$}
\eea
It is possible to have $C_\mathcal{B}=\un{u}$ iff $KG_{\kappa}(:,\mathcal{B})=\un{u}-\un{v}G_s(:,\mathcal{B})$. There exists a solution to $K$ for every $\un{u}, \un{v}$ iff $G_{\kappa}(:, \mathcal{B})$ is of rank $j$. It is clear to see that secrecy fails when  $G_{\kappa}(:, \mathcal{B})$ is not a rank $j$ matrix, as the numerator term in equation~\eqref{eq:bayes} could be zero for some choice of $\un{u}$ and  $\un{v}$. When \scalebox{0.92}{$G_{\kappa}(:,\mathcal{B})$ }
has full rank, there are exactly $q^{k-\hat{k}-j}$ solutions for $K$. This results in  $ Pr(C_\mathcal{B}= \un{u} \mid S= \un{v})
={q^{-j}}$ for all $\un{u}, \un{v}$. Substituting the above equation into equation~\eqref{eq:bayes}, we get the condition required for secrecy, proving the lemma.
\eprf

We now present the definition of nested matrix used in \cite{zhang_secure_25} to construct secure convertible codes and claim that the construction in \cite{zhang_secure_25} is secure under the scenarios where the eavesdropper accesses $\ell$ unchanged symbols, but could fail under other general scenarios where the eavesdropper accesses a mix of unchanged, retired, and new symbols.

\bdefn[Nested Matrix]\label{def:nestedcode}  A $k \times n$ matrix of the form $G$ is said to be $[n,k]$ $\ell$-nested if (a) $G$ is a generator matrix of an $[n,k]$ MDS code and (b) given $G=[G_s^T~G_{\kappa}^T]^T$ where $G_s$ is  a $(\hat{k} \times n)$ matrix and $G_{\kappa}$ is a $((k-\hat{k}) \times n )$ matrix, $G_{\kappa}$ is a generator matrix of an $[n, k-\hat{k}]$ MDS code. 
\edefn

\bcor \label{cor:nestedcode}
An $[n, k]$ MDS code $\mathcal{C}$ is $(\ell, k-\ell)$-secure iff there exists a generator matrix $G$ of $\mathcal{C}$ that is $[n,k]$  $\ell$-nested. 
\ecor
\bprf
Proof follows from Lemma~\ref{lem:secrecycondition}. 
\eprf








\subsection{Nested Matrix-Based Precoding}

We describe the secure code construction from \cite{zhang_secure_25} for the case when $\lambdai \ge \lambdaf$. The initial secure code is described through the generator matrix $G^I = P \hat{G}^I$, where $\hat{G}^I$ is the generator matrix of the initial code of an access-optimal MDS convertible code and $P$ is the precoding matrix of the form:
\bean
P = \left[ \begin{array}{c|c}
    I_{k^I-\ell} & \bold{0} \\ \hline 
    \multicolumn{2}{c}{P_{\kappa}}
\end{array} \right]  
\eean
where $P_{\kappa}$ is chosen to be generator matrix of any $[k^I, \ell]$ MDS code ensuring that $P$ is an $[k^I, k^I]$ $\ell$-nested matrix. If $\hat{G}^I$ is a systematic generator matrix, this directly implies that if an eavesdropper accesses any $\ell$ symbols out of the first $k^I$ symbols of the codeword, they obtain no information about the secret message symbol due to the nested code properties. However, if an eavesdropper accesses any $\ell$ symbols from the codeword, it is not enough that $P$ is an $[k^I, k^I]$ $\ell$-nested matrix, it is also required that the generator matrix $G^I$ is an $[n^I, k^I]$ $\ell$-nested matrix to guarantee $\ell$-secrecy by Corollary~\ref{cor:nestedcode}.

\begin{example}
Let us consider an example where $[n^I, k^I] = [3, 2]$. Let
$P =
\begin{bmatrix}
1 & 0 \\
1 & 1
\end{bmatrix}, 
\hat{G}^I =
\begin{bmatrix}
1 & 0 & 1 \\
0 & 1 & 1
\end{bmatrix}.$ It is evident that $P$ is a $[2,2]$ $1$-nested matrix. The encoded vector can be expressed as
$[s~\kappa] P \hat{G}^I = [{s} + \kappa ~ \kappa] \hat{G}^I = [s+\kappa~\kappa~s]$.
It is evident that the last symbol directly reveals the secure message symbol ${s}$. Therefore, an eavesdropper observing even a single symbol can recover $s$, violating the secrecy requirement.
\end{example}

This example demonstrates that the masking structure introduced by nested matrix-based precoding guarantees secrecy when the symbols accessed by the eavesdropper are restricted to the message/unchanged symbols. However, it is not guaranteed to be secure when the eavesdropper accesses parity symbols. In this paper, we construct secure convertible codes that remain secure when the eavesdropper accesses any $\ell$ symbols across the initial and final codewords.


\section{Our Secure Convertible Codes Framework \label{sec:secureframework}}
In this section, we present a framework for constructing secure convertible codes. Let $[n^I,k^I]$ and $[n^F,k^F]$ denote the parameters of the initial and final codes, respectively. We consider the \emph{general regime} of convertible codes, where $\lambdai$ initial codewords are converted to $\lambdaf$ final codewords. First we describe the constructions in the regime where $\lambdai \ge \lambdaf$. The constructions in the regime $\lambdai \le \lambdaf$ follow exactly in similar lines with the initial and final code roles exchanged as shown in Section~\ref{sec:othercase}. 

We use existing constructions of access-optimal convertible codes in the literature as building blocks for the secure convertible code constructions. Let $\hat{G}^I \in \mathbb{F}_q^{k^I \times n^I} $and $\hat G^F\in\mathbb{F}_q^{k^F\times n^F}$ denote the generator matrices of the initial and final codes of an access-optimal MDS convertible code which need not be secure. Constructions of such codes can be found in \cite{Maturana_Rashmi_TIT, Kong_LRC_TIT, Convertible_ISIT25}. The framework we describe allows for the description of initial and final codes of secure convertible code defined over an extension field $\mathbb{F}_{q^m}$ provided that the generator matrices $\hat{G}^I$ and $\hat{G}^F$ have entries from $\mathbb{F}_q$. In Section~\ref{sec:mrd} MRD based constructions require $m \ge \min(k^I, k^F)$ whereas in Section~\ref{sec:grs} GRS based constructions have $m=1$.

\subsection{$\lambdai \ge \lambdaf$}
In this regime where $\lambdai \ge \lambdaf$, the secrecy capacity upper bound is given by $k_S \le \lambdai(k^I-\ell)$. The convertible code is described through a unified generator matrix later in this section that helps with the proof of $\ell$-secrecy.
\subsubsection{Initial Code}
We set $k_S = \lambdai(k^I-\ell)$.
For each $i\in[\lambdai]$, let $\underline{s}^{(i)}\in\mathbb{F}_{q^m}^{k^I-\ell}$ denote the secure message vector and let $\underline{\kappa}^{(i)}\in\mathbb{F}_{q^m}^{\ell}$ denote the key vector. These are concatenated to form the message vector 
$ \underline{m}^{(i)}=[\underline{s}^{(i)}\;\;\underline{\kappa}^{(i)}]$ and precoded by the invertible precoder matrix $P$ to obtain  
$
\underline{\tilde m}^{(i)} = \underline{m}^{(i)}P \text{ where } P=
\begin{bmatrix}
P_s^\top&
P_{\kappa}^\top
\end{bmatrix}^\top
\in\mathbb{F}_{q^m}^{k^I\times k^I},
$ with  $P_s \in\mathbb{F}_{q^m}^{(k^I-\ell) \times k^I},
P_{\kappa} \in\mathbb{F}_{q^m}^{\ell\times k^I}.$ 
The resulting vector $ \un{\tilde m}^{(i)}$ is then encoded by the initial convertible code generator $\hat G^I$, yielding the initial codeword $
\underline{c}^{(i)}
=
\underline{\tilde m}^{(i)}\hat G^I
=
\un{m}^{(i)}P\hat G^I.$ 
Note that $G^I=P\hat{G}^I$ forms a generator matrix of an $[n^I, k^I]$ MDS code defined over $\mathbb{F}_{q^m}$ as long as $P$ is invertible.

\subsubsection{Unified Generator Matrix} In order to prove the $\ell$-secrecy property of the convertible code, we need to consider cases where a subset of initial codeword symbols and a subset of final codeword symbols are accessed by the eavesdropper. We therefore describe the unified generator matrix $G_{\text{joint}}$ of size $(\lambdai k^I \times (\lambdai n^I + \lambdaf n^F))$ that maps the  $\lambdai k^I$ length message vector $[\un{s}~\un{\kappa}]$ to a codeword of length $\lambdai n^I + \lambdaf n^F$ as shown below: 
\bean
\label{eq:jointenc}
[\un{c}^{(1)}~\cdots~\un{c}^{(\lambdai)}\un{d}^{(1)}~\cdots~\un{d}^{(\lambdaf)}] = [\un{s}~\un{\kappa}] G_{\text{joint}},
\eean
where $\un{s}=[\un{s}^{(1)}\cdots\un{s}^{(\lambdai)}]$,
$\un{\kappa}=[\un{\kappa}^{(1)}\cdots\un{\kappa}^{(\lambdai)}]$, and $\un{d}^{(j)}$ is the $j$-th final codeword. The joint generator matrix that collectively describes the encoding of $\lambdai k^I$ message symbols to all initial and final codeword symbols is of the form $G_{\text{joint}} = P_D \hat{G}_{\text{joint}}$. Here  $\hat{G}_{\text{joint}}$ is of same size as ${G}_{\text{joint}}$ and  
\bean
P_D=\begin{bmatrix}
  P_{s, D}\\
  P_{\kappa, D}
  \end{bmatrix} , P_{u, D} = \mathrm{diag}(P_u,\dots,P_u)
\eean
  for  $u \in \{s, \kappa \}$  and  $ P_{s, D} \in\mathbb{F}_{q^m}^{\lambdai (k^I-\ell)\times  \lambdai k^I}$,   $P_{\kappa, D}\in\mathbb{F}_{q^m}^{\lambdai \ell \times\lambdai k^I}$ and $\hat{G}_{\text{joint}}$ is defined as shown below:
\bean
\label{eq:Gjoint}
\scalebox{0.8}{$\hat{G}_{\text{joint}} = \left[ \begin{array}{cccc|cccc}
     \hat{G}^I & 0 & \cdots & 0 & \multirow{4}{*}{$\hat{G}^F_{\text{ext}, 1}$}& \multirow{4}{*}{$\hat{G}^F_{\text{ext}, 2}$}&\multirow{4}{*}{$\cdots$} & \multirow{4}{*}{$\hat{G}^F_{\text{ext}, \lambdaf}$}\\
     0 & \hat{G}^I & \cdots & 0 &  \\
     \vdots & \vdots & \ddots & 0 &    \\
     0 & 0 & \cdots &  \hat{G}^I &  \\
\end{array}\right]$}.
\eean
It can be verified that the first $\lambda^I n^I$ columns defined in $\hat{G}_{\text{joint}}$ imply $\un{c}^{(i)} = [\un{s}^{(i)}~\kappa^{(i)}] P \hat{G}^I$. We provide the exact description of the extended matrices $\hat
{G}_{\text{ext}, j}$ in equations~\eqref{eq:Gextj} and \eqref{eq:Gextji}.  

\blem\label{lem:Gjointsecrecycondition}
The $[n^I, k^I; n^F, k^F]$ convertible code described by the joint generator matrix $G_{\text{joint}} = P_D \hat{G}_{\text{joint}}$ satisfies $\ell$-secrecy if the matrix $P_{\kappa, D} \hat{G}_{\text{joint}}(:, \mathcal{B})$ has full column rank for every $\mathcal{B} \subseteq [\lambdai n^I +\lambdaf n^F]$ such that $|\mathcal{B}| \le \ell$ and $\hat{G}_{\text{joint}}(:, \mathcal{B})$ is of full rank.
\elem
\bprf
We first note that $G_{\text{joint}}(:, \mathcal{B})$ is of full rank iff $\hat{G}_{\text{joint}}(:, \mathcal{B})$ is of full rank. This follows as $P_D$ is an invertible matrix. The rest of the proof follows from Lemma~\ref{lem:secrecycondition}.
\eprf

\subsubsection{Final Code}\label{subsec:final} Recall the initial and final partitions $\{\mathcal{U}_i\}_{i\in[\lambdai]}$ and
$\{\mathcal{V}_j\}_{j\in[\lambdaf]}$
defined in Section~\ref{intro}. W.l.o.g. we assume that $\mathcal{U}_i = [(i-1)k^I+1:ik^I]$. For each final partition $\mathcal{V}_j$, define $\mathcal{S}_j=\{\,i\in[\lambdai]:\mathcal{U}_i\cap \mathcal{V}_j\neq\emptyset\,\}$ as the collection of initial codewords from which message symbols of $j$-th codeword are picked, and let
$\mathcal{S}_j=\{j_1,j_2,\ldots,j_{{s}_j}\},$ where ${s}_j=|\mathcal{S}_j|$.  Further, define 
$
\mathcal{V}_{j,i}= \{ u-(i-1)k^I \mid u \in  \mathcal{U}_i\cap \mathcal{V}_j \} 
$
for all $i\in \mathcal{S}_j$, and it can be see that $k^F=\sum_{i\in \mathcal{S}_j}|\mathcal{V}_{j,i}|$. The $j$-th final message vector is obtained by concatenating the symbols indexed by the sets $\{\mathcal{V}_{j, i}\}_{i\in \mathcal{S}_j}$, i.e.,
\bean
\label{eq:mfinalj}
\underline{\tilde{m}}_F^{(j)}
=
\Big[
\underline{\tilde{m}}^{(j_1)}(\mathcal{V}_{j,j_1})
~
\underline{\tilde{m}}^{(j_2)}(\mathcal{V}_{j,j_2})
~
\cdots~
\underline{\tilde{m}}^{(j_{{s}_j})}(\mathcal{V}_{j,j_{{s}_j}})
\Big].
\eean

The $j$-th final codeword is given by $
\underline{d}^{(j)}
= \underline{\tilde{m}}_F^{(j)}\hat{G}^F.$
We will now define an extended generator matrix ${G}_{\text{ext}, j}^F$, an $\lambdai k^I \times n^F$ matrix that represents the encoding of vector  $[\un{s}, \un{\kappa}]$ to a $j$-th final codeword containing  $n^F$ symbols. Here, $\un{s}=[\un{s}^{(1)}\cdots\un{s}^{(\lambdai)}]$,
$\un{\kappa}=[\un{\kappa}^{(1)}\cdots\un{\kappa}^{(\lambdai)}]$.
Let $G_{ext, j}^F = P_{D} \hat{G}_{ext, j}^F$ and

\bea
\label{eq:Gextj}
\hat{G}_{ext, j}^F = \begin{bmatrix}
     {\hat{G}_{ext, j, 1}^F}{}^\top &
    {\hat{G}_{ext, j, 2}^F}{}^\top & 
     \cdots &    {\hat{G}_{ext, j, \lambdai}^F}{}^\top
 \end{bmatrix}^\top,
\eea

where $\hat{G}_{ext, j, i}^F$ is an $k^I \times n^F$ matrix for every $j \in [\lambdaf]$ and $i \in [\lambdai]$, 
The $j$-th final codeword depends only on message symbols of initial codewords whose index is in $S_j$. Therefore, it is clear to see that $\hat{G}_{ext, j, i}^F=0$ for $i \notin \mathcal{S}_j$. The remaining matrices are defined as follows:
\bea
\label{eq:Gextji}
\hat{G}_{\text{ext}, j, j_i}^F(u, :) = \begin{cases}
    \hat{G}^F(m_i+p_{i,j}(u), :) & u \in \mathcal{V}_{j,j_i} \\
    0 & \text{otherwise}.
\end{cases}
\eea
for $i \in [s_j]$ where $m_i=\sum_{\hat{i}=1}^{i -1}|\mathcal{V}_{j, j_{\hat{i}}}|$ and $p_{i, j}(u)$ is the position of $u$ in $\mathcal{V}_{j, j_i}$ given its entries are sorted in ascending order.
Using the extended generator matrix, it can be verified that
$\un{d}^{(j)} = \underline{\tilde{m}}_F^{(j)}\hat{G}^F =  [\un{s}~\un{\kappa}] G_{\text{ext}, j}^F =  [\un{s}~\un{\kappa}] P_D \hat{G}_{\text{ext}, j}^F.$
In merge regime, it can be seen that $\hat{G}^F_{\text{ext}, j} = \hat{G}^F$ as  $|\mathcal{S}_1|=\lambdai$ and $|\mathcal{V}_{1,i}|=k^I$ for all $i \in [\lambdai]$. 

\begin{example}
Let $[n^I=4, k^I=2; n^F=5, k^F=3]$ as shown in Fig.~\ref{fig:convcode} implying that $\lambdai=3$, $\lambdaf=2$.
Take $\mathcal{U}_1=\{1,2\}$, $\mathcal{U}_2=\{3,4\}$, $\mathcal{U}_3=\{5,6\}$. Then 
\bean
\mathcal{V}_1=\{1,2,3\} \implies \mathcal{S}_1=\{1,2\}, \mathcal{V}_{1,1}=\{1,2\}, \mathcal{V}_{1,2}=\{1\}, \\
\mathcal{V}_2=\{4,5,6\} \implies \mathcal{S}_2=\{2,3\}, \mathcal{V}_{2,2}=\{2\}, \mathcal{V}_{2,3}=\{1,2\}.
\eean


Each $\hat G^F_{\mathrm{ext},j}\in\mathbb{F}_q^{6\times 5}$ consists of
$\lambdai=3$ blocks $\hat G^F_{\mathrm{ext},j,i}$ of two rows each. By equation~\eqref{eq:Gextji},
\bean
\scalebox{0.9}{$\hat{G}_{\mathrm{ext},1}^F$}&=&\scalebox{0.9}{$
\begin{bmatrix}
\hat G^F(1,:)^\top&
\hat G^F(2,:)^\top&
\hat G^F(3,:)^\top&
\mathbf{0}& 
\mathbf{0}&
\mathbf{0}
\end{bmatrix}^\top$}
\\
\scalebox{0.9}{$\hat{G}_{\mathrm{ext},2}^F$}&=&
\scalebox{0.9}{$\begin{bmatrix}
\mathbf{0}&
\mathbf{0}& 
\mathbf{0}&
\hat G^F(1,:)^\top& 
\hat G^F(2,:)^\top & 
\hat G^F(3,:)^\top 
\end{bmatrix}^\top $}
\eean
where each $\mathbf{0}$ is a $5\times 1$ zero column. 
\end{example}

\bnote \normalfont \label{rem:decod} In this framework, the matrices $P, \hat{G}^I, \hat{G}^F$ completely describe a secure convertible code, and the access cost of it remains the same as the access cost of a convertible code whose initial and final codes are described through $\hat{G}^I, \hat{G}^F$ respectively. The decoding property holds true in the merge regime, as $\hat{\un{m}}^{(i)}$ can be recovered from any $k^I$  symbols of $i$-th codeword, and $\un{s}^{(i)}$ can be recovered from it as $P$ is invertible. $\hat{\un{m}}_F^{(1)}$ can be recovered from any $k^F=\lambdai k^I$  symbols of the final codeword and $\un{s}$ can be recovered from it as $P_D$ is invertible.  The decoding property for the split regime can also be argued similarly.
\enote

\subsection{$\lambdai < \lambdaf$}\label{sec:othercase}
We consider the general case $\lambda^I<\lambda^F$ and secrecy
capacity bound is $k_S=\lambda^F(k^F-\ell).$ For each $j\in[\lambda^F]$, let
$\underline{s}^{(j)}\in\mathbb{F}_{q^m}^{k^F-\ell}$ and
$\underline{\kappa}^{(j)}\in\mathbb{F}_{q^m}^{\ell}$
denote the secure-message and key vectors, respectively. Define
\bean
\underline{s}
=[\underline{s}^{(1)}\;\cdots\;
\underline{s}^{(\lambda^F)}],
\qquad
\underline{\kappa}
=[\underline{\kappa}^{(1)}\;\cdots\;
\underline{\kappa}^{(\lambda^F)}].
\eean
\subsubsection{Final Secure Code}

 The message vector $\underline{m}^{(j)}
=
[\underline{s}^{(j)}\;\underline{\kappa}^{(j)}]
\in\mathbb{F}_{q^m}^{k^F}$ is first encoded using an invertible precoder matrix
\bean
P=
\begin{bmatrix}
P_s\\
P_{\kappa}
\end{bmatrix}
\in\mathbb{F}_{q^m}^{k^F\times k^F},
\eean
where $P_s\in\mathbb{F}_{q^m}^{(k^F-\ell)\times k^F}, P_{\kappa}\in\mathbb{F}_{q^m}^{\ell\times k^F}.$
The resulting vector $\underline{\tilde{m}}^{(j)}
=
\underline{m}^{(j)}P$ is then encoded using the final convertible-code generator matrix $\hat{G}^F$, yielding the final codeword $\underline{d}^{(j)}
=
\underline{\tilde{m}}^{(j)}\hat{G}^F
=
\underline{m}^{(j)}P\hat{G}^F.$ Therefore, the effective generator matrix in the final configuration is $G^F=P\hat{G}^F.$

\subsubsection{Initial Code}
\label{subsec:initial}

The initial and final coordinate partitions are
$\{\mathcal{U}_i\}_{i\in[\lambda^I]}$ and
$\{\mathcal{V}_j\}_{j\in[\lambda^F]}$, respectively.
Without loss of generality, assume
$\mathcal{V}_j=[(j-1)k^F+1:jk^F]$.
For each initial partition $\mathcal{U}_i$, define $\mathcal{S}_i
=\{j\in[\lambda^F]:
\mathcal{U}_i\cap\mathcal{V}_j\neq\emptyset\},$ which identifies the final codewords whose message symbols
contribute to the $i$-th initial codeword. Further, define
\bean
\mathcal{U}_{i,j}
=\{u-(j-1)k^F:
u\in\mathcal{U}_i\cap\mathcal{V}_j\},
\eean
for each $j\in\mathcal{S}_i$. Let $\mathcal{S}_i=\{i_1,i_2,\ldots,i_{{s}_i}\},$ where ${s}_i=|\mathcal{S}_i|$. The $i$-th initial message vector
is obtained by concatenating the symbols indexed by
$\{\mathcal{U}_{i,j}\}_{j\in\mathcal{S}_i}$ in increasing
order of $j$. Since these sets partition the coordinates of
$\mathcal{U}_i$, their cardinalities satisfy
$k^I=\sum_{j\in\mathcal{S}_i}|\mathcal{U}_{i,j}|.$
The $i$-th initial codeword is given by
$\underline{c}^{(i)}
=\underline{\tilde m}_I^{(i)}\hat{G}^I,$
where $\underline{\tilde m}_I^{(i)}$ denotes the
$i$-th initial message vector. 
\bean
\label{eq:minitialj}
\underline{\tilde{m}}_I^{(i)}
=
\Big[
\underline{\tilde{m}}^{(i_1)}(\mathcal{U}_{i,i_1})
~
\underline{\tilde{m}}^{(i_2)}(\mathcal{U}_{i,i_2})
~
\cdots~
\underline{\tilde{m}}^{(i_{{s}_i})}(\mathcal{V}_{i,i_{{s}_i}})
\Big].
\eean We define the extended initial
generator matrix $G^I_{\mathrm{ext},i}$, of dimensions
$\lambda^F k^F\times n^I$, to represent the encoding of
$[\underline{s}\ \underline{\kappa}]$ into the $i$-th
initial codeword. Let $G^I_{\mathrm{ext},i}
=P_D\hat{G}^I_{\mathrm{ext},i},$ where  $P_D=\begin{bmatrix}
  P_{s, D}^T~
  P_{\kappa, D}^T
  \end{bmatrix}^T
,$ such that
\bean
P_{u,D}=\operatorname{diag}
(\underbrace{P_u,\ldots,P_u}_{\lambda^F\text{ blocks}}) \text{ for } u \in \{s, \kappa\} \text{ and },
\eean
\bean
\hat{G}^I_{\mathrm{ext},i}
=
\begin{bmatrix}
(\hat{G}^I_{\mathrm{ext},i,1})^T&
\cdots&
(\hat{G}^I_{\mathrm{ext},i,\lambda^F})^T
\end{bmatrix}^{T}.
\eean 
Here, each $\hat{G}^I_{\mathrm{ext},i,j}$ is a
$k^F\times n^I$ matrix. For $j\in\mathcal{S}_i$, its rows
are defined by

\bean
\hat{G}^I_{\mathrm{ext},i,j}(v,:)
=
\begin{cases}
\hat{G}^I(p_i(u),:),
& u=(j-1)k^F+v\in\mathcal{U}_i,\\
\underline{0},&\text{otherwise},
\end{cases}
\eean
where $p_i(u)$ denotes the position of $u$ in $\mathcal{U}_i$,
whose elements are ordered increasingly, and
$v\in[k^F]$. For $j\notin\mathcal{S}_i$,
$\hat{G}^I_{\mathrm{ext},i,j}=\mathbf{0}$.

Using this extended generator matrix, the $i$-th initial
codeword can be written as $\underline{c}^{(i)}
=[\underline{s}\ \underline{\kappa}]
G^I_{\mathrm{ext},i}
=[\underline{s}\ \underline{\kappa}]
P_D\hat{G}^I_{\mathrm{ext},i}.$

\subsubsection{Unified Generator Matrix}
\label{subsec:jointenc_less}

To establish the $\ell$-secrecy property of the convertible
code for $\lambda^I<\lambda^F$, we consider an eavesdropper
accessing symbols from both the initial and final codewords.
We define the joint generator matrix $G_{\mathrm{joint}}$ of
size $\lambda^F k^F\times
(\lambda^I n^I+\lambda^F n^F)$, which jointly describes the
encoding of all initial and final codewords. We assume
$\lambda^I k^I=\lambda^F k^F$.

Let $P\in\mathbb{F}_{q^m}^{k^F\times k^F}$ be an invertible
precoding matrix, partitioned as $P=\begin{bmatrix}
P_s^T~
P_\kappa^T
\end{bmatrix}^T,$ where $P_s$ has $k^F-\ell$ rows and $P_\kappa$ has $\ell$
rows. 

The joint generator matrix is given by $G_{\mathrm{joint}}=P_D\hat{G}_{\mathrm{joint}},$ where $\hat{G}_{\mathrm{joint}}
=
\left[
\hat{G}^{I}_{\mathrm{ext},1}\;\cdots\;
\hat{G}^{I}_{\mathrm{ext},\lambda^I}
\;\middle|\;
\operatorname{diag}
(\underbrace{\hat{G}^{F},\ldots,
\hat{G}^{F}}_{\lambda^F\text{ blocks}})
\right]$ and 
 $P_D=\begin{bmatrix}
  P_{s, D}^T~
  P_{\kappa, D}^T
  \end{bmatrix}^T
.$
Here, $\hat{G}^{I}_{\mathrm{ext},i}$ is the extended
generator matrix that maps the concatenated final-message vector
to the $i$-th initial codeword, as defined in
Section~\ref{subsec:initial}. The extended final generator
matrices are defined in Section~\ref{subsec:final}.
Consequently, $\hat{G}_{\mathrm{joint}}
\in\mathbb{F}_{q^m}^{\lambda^F k^F
\times(\lambda^I n^I+\lambda^F n^F)}.$

\blem\label{lem:Gjointsecrecycondition_general}
The convertible code described by
$G_{\mathrm{joint}}=P_D\hat{G}_{\mathrm{joint}}$
satisfies $\ell$-secrecy if, for every
$\mathcal{B}\subseteq
[\lambda^I n^I+\lambda^F n^F]$ such that
$|\mathcal{B}|\leq\ell$ and
    $\hat{G}_{\mathrm{joint}}(:,\mathcal{B})$ has full column
rank, the key-component matrix
$P_{\kappa,D}\hat{G}_{\mathrm{joint}}(:,\mathcal{B})$
also has full column rank.
\elem

\bprf
Since $P_D$ is invertible,
$G_{\mathrm{joint}}(:,\mathcal{B})$ and
$\hat{G}_{\mathrm{joint}}(:,\mathcal{B})$ have the same
rank. The result follows from
Lemma~\ref{lem:secrecycondition}.
\eprf



\section{Secure Convertible Codes from MRD Codes \label{sec:mrd}}
In this section, we present constructions of secure convertible codes using Maximum Rank Distance (MRD) codes. 
Pick an extension an extension field  $\mathbb{F}_{q^m}$ of $\mathbb{F}_q$. Let
$\mathcal{C}\subseteq\mathbb{F}_{q^m}^{\,n}$ be a linear code of dimension $k$. For $\underline{c}=(c_1,\ldots,c_n)\in\mathbb{F}_{q^m}^n$, the rank of
$\underline{c}$ is defined as
$\operatorname{rank}_q(\underline{c})
=
\operatorname{dim}_{\mathbb{F}_q}
\left(
\langle c_1,c_2,\ldots,c_n\rangle
\right)$.
The
\emph{rank distance} between two codewords
$\underline{c},\underline{c}'\in\mathcal{C}$ is defined as $d_R(\underline{c},\underline{c}')
=
\operatorname{rank}_q(\underline{c}-\underline{c}').$ The minimum rank distance of $\mathcal{C}$ is
$
d
=
\min_{\substack{
\underline{c} \ne \underline{c}'\in\mathcal{C}
}}
d_R(\underline{c},\underline{c}').
$
The Singleton bound for
rank-metric states that $d\le n-k+1$, for $n\le m,$ see, e.g.,~\cite[Sec.~2]{Bartz2022}. Codes attaining the Singleton bound with equality are called \emph{Maximum Rank Distance (MRD) codes}. Gabidulin codes constitute an important class of MRD codes~\cite{Gabidulin1985}.
 See \cite{Bartz2022} for an overview of MRD codes.


\textit{MRD based precoding matrix:}
We consider the regime where $\lambdai \ge \lambdaf$ and describe the precoding matrix through the generator matrix of an MRD code as shown below.
Choose $P = \left[\begin{array}{c|c}
    I_{k^I -\ell} & \bold{0} \\ \hline
    \multicolumn{2}{c}{P_{\kappa}}
\end{array} \right]$ where $P_{\kappa}$ is a generator matrix of an $[k^I, \ell]$ MRD code defined over $\mathbb{F}_{q^m}$. In Theorem~\ref{thm:MRD}, we show that this choice of precoding matrix satisfies the sufficiency conditions needed for the $[n^I, k^I; n^F, k^F]$ convertible code to satisfy $\ell$-secrecy as shown in Lemma~\ref{lem:Gjointsecrecycondition}.
Known MRD constructions like Gabidulin codes require $m \ge k^I$ to construct an $[k^I, \ell]$ MRD code. This implies $[n^I, k^I; n^F, k^F]$ convertible code constructed using MRD code requires a field size $O(q^{k^I})$ in the regime, where $q$ is the field size required for access-optimal convertible code with parameters $[n^I, k^I; n^F, k^F]$.

\bthm \label{thm:MRD} Let $\lambdai \ge \lambdaf$, $P = \left[\begin{array}{c|c}
    I_{k^I -\ell} & \bold{0} \\ \hline
    \multicolumn{2}{c}{P_{\kappa}}
\end{array} \right]$ where $P_{\kappa}$ is a generator matrix of an $[k^I,\ell]$ MRD code, and let $\hat{G}^I$, $\hat{G}^F$ be systematic generator matrices of initial and final codes of any $[n^I, k^I; n^F, k^F]$ access-optimal convertible code. The convertible code described through the matrices $P, \hat{G}^I, \hat{G}^F$ satisfies $\ell$-secrecy  property. 
\ethm
\bprf 
We show that if $\hat{G}_{\text{joint}}(:, \mathcal{B})$ is full rank for a $B \subseteq [\lambdai n^I + \lambdaf n^F]$, then $P_{\kappa, D} \hat{G}_{\text{joint}}(:, \mathcal{B})$ is a full rank matrix and from Lemma~\ref{lem:Gjointsecrecycondition} the $\ell$-secrecy property follows.
Let $\mathcal{B}\subseteq[n^I]$ with $|\mathcal{B}|= j \le \ell$, and let $M=P_{\kappa, D} \hat{G}_{\text{joint}}(:, \mathcal{B})$. Suppose $M$ is not full rank. Then, there exists a nonzero vector $
\underline{v}=[v_1,v_2,\ldots,v_j]^T
\in\mathbb{F}_{q^m}^{j}$ such that $M\underline{v}=0.$ Let $\underline{y}=\hat{G}_{\text{joint}}(:, \mathcal{B})\underline{v}.$ Then $P_{\kappa, D}\underline{y}=0.$ Since $\hat{G}_{\text{joint}}(:, \mathcal{B})$ is full rank, it follows that $\un{y} \ne 0$ if $\un{v} \ne  0$. Let $\un{y}=(\un{y}_1^T, \un{y}_2^T, \cdots, \un{y}_{\lambdai}^T)^T$ where $\un{y}_i^T \in \mathbb{F}_{q^m}^{k^I}$. $
P_{\kappa, D}\underline{y}=0$ implies that $ P_{\kappa} \underline{y}_i = 0 ~ \forall i \in [\lambdai]$. Since $\un{y} \ne 0$ it implies that there exists an $i \in [\lambdai]$ such that $ \underline{y}_i \ne 0$.  Since $P_{\kappa}$ is the generator matrix of an $[k^I,\ell]$ MRD code
$\mathcal{C}$ and $P_{\kappa}\underline{y}_i=0,$ it follows that $\underline{y}_i\in\mathcal{C}^{\perp}.$ Since the dual of an MRD code is again an MRD code, $\mathcal{C}^{\perp}$
is an $[k^I,k^I-\ell]$ MRD code. Therefore, its minimum rank distance is $d^{\perp}
=
k^I-(k^I-\ell)+1
=
\ell+1.
$ As $\underline{y}_i\neq0$ is a codeword of $\mathcal{C}^{\perp}$, $\operatorname{rank}_{\mathbb{F}_q}(\underline{y}_i)
\ge
\ell+1.$ On the other hand, 
since every entry of $\hat{G}_{\text{joint}}$ belongs to $\mathbb{F}_q$, each
entry of $\un{y}_i$ is an $\mathbb{F}_q$-linear combination of
$v_1,v_2,\ldots,v_\ell$. Hence $\operatorname{rank}_{\mathbb{F}_q}(\underline{y}_i) \le \ell$, resulting in a contradiction and proving the theorem.
\eprf

\bnote \normalfont \label{rem:MRDdecod}
For the case $\lambdaf \geq \lambdai$, the $\ell$-secrecy proof follows similarly by interchanging the roles of the initial and final configurations (see Section~\ref{sec:othercase}). As argued in Remark~\ref{rem:decod}, the decoding property holds in merge and split regimes. Therefore, our MRD based construction results in access-optimal secure convertible codes in merge and split regimes with field size $O(q^{\min(k^F, k^I)})$, where $q$ is the field size required for the underlying access-optimal convertible code.
\enote

\section{Secure Convertible Codes from GRS codes \label{sec:grs}}

The MRD-based construction presented in the previous section guarantees the required secrecy properties; however, the field size grows exponentially with $\min(k^I, k^F)$, leading to increased implementation complexity. In this section, we present an alternative construction based on Generalized Reed-Solomon (GRS) codes that achieves the same secrecy guarantees while requiring only a base field of size $q$ for the merge and split regimes. Thus, the proposed GRS-based construction significantly reduces the field size requirement from exponential to linear in the code length.

\bdefn[GRS code]

An \scalebox{0.92}{$[n,k,\underline{u},\mathcal{E}]_q$} Generalized Reed-Solomon (GRS) code
$\mathcal{C}$ over $\mathbb{F}_q$ is defined as
\bean
\scalebox{0.9}{$
\mathcal{C}
=
\left\{
\left(
u_{1}f(\alpha_1),
u_{2}f(\alpha_2),
\ldots,
u_{n}f(\alpha_n)
\right)
\,\middle|\,
f\in\mathbb{F}_q[x],\;
\deg(f)<k
\right\}
$}.
\eean
where $\und{u}
=
(u_{1},u_{2},\ldots,u_{n})$ is a vector of $n$ nonzero elements of $\mathbb{F}_q$, and $\mathcal{E}=\{\alpha_1,\alpha_2,\ldots,\alpha_n\}\subseteq\mathbb{F}_q$ is a set of $n$ distinct evaluation points.
\edefn


We provide a construction of secure convertible codes in the merge regime when $\lambdai \le \max(r^F, r^I)$, where the initial and final codes of the access-optimal convertible code are from the GRS family. Our proof of the $\ell$-secrecy property requires that the generator matrices of the initial and final codes are constructed using a parallel-block-reconstructible matrix. 

\bdefn[Parallel-block-reconstructible matrix \cite{Myna_TIT_MDS} ]\label{def:pbr}
We say that a super-regular matrix $M \in \mathbb{F}_q^{\lambdai k^I\times r}$ is \emph{parallel-block-reconstructible} if, for every $j\in[\lambdai]$ and every $i\in[r]$, there exists an index $i_j\in[r]$ and a nonzero scalar $c_{i,j} \in\mathbb{F}_q$ such that $M([(j-1)k^I+1:jk^I],i) = c_{i,j}\, {M}([k^I],i_j)$. Let $r^I = n^I-k^I$ and  $r^F=n^F-k^F$. \edefn

Given a $\lambdai k^I \times \max(r^I, r^F)$ parallel-block-reconstructible matrix $M$, let  
\bea
\label{eq:genGRS}
\hat{G}^I = [I\quad M([k^I],[r^I])], \hat{G}^F = [I\quad M(:,[r^F])].
\eea
Convertible codes with initial and final code generator matrices as $\hat{G}^I$ and $\hat{G}^F$ (see equation~\eqref{eq:genGRS}) are access-optimal $[n^I, k^I; n^F, k^F=\lambda k^I]$ MDS convertible codes. Constructions of parallel-block-reconstructible matrices are known for all parameters in \cite{Maturana_Rashmi_TIT}, but they require large field sizes and do not result in initial and final codes that are GRS. In \cite{Myna_TIT_MDS}, linear field size parallel-block-reconstructible matrices are constructed for any $\lambdai \le \max(r^F,r^I)$ that result in final and initial codes being GRS codes. We will use this construction to obtain secure convertible codes.

Let $\hat{G}^I, \hat{G}^F$ be systematic generator matrices of GRS codes described through equation~\eqref{eq:genGRS}. Let $V_d(\{\alpha_1, \cdots, \alpha_n\})$ be the $d \times n$ matrix whose $(i, j)$-th entry is $\alpha_j^{i-1}$. 
Given the initial code described through $\hat{G}^I$ is an $[n^I, k^I, \un{u}, \mathcal{X} \cup \mathcal{Y}]$ GRS code,  
we pick  the precoder matrix 
\bea
\label{eq:grs_precoder}
P=\mathrm{V}_{k^I}(\mathcal{X}) \text{diag}(u_1, u_2, \cdots, u_{k^I}),
\eea
where, $\mathcal{X}$, $\mathcal{Y}$ be two disjoint sets of $\mathbb{F}_q$ with  $k^I, r^I$ elements respectively. 

\bthm\label{thm:GRSsecureconv}
 The $[n^I=k^I+r^I, k^I; n^F=k^F+r^F, k^F = \lambdai k^I ]$ convertible code described through the matrices $P, \hat{G}^I, \hat{G^F}$ (shown in equations~\eqref{eq:grs_precoder},\eqref{eq:genGRS}) for $\lambdai \le \max(r^I, r^F)$ is an $(\ell, k_S)$-secure convertible code with $k_S=\lambdai(k^I-\ell)$.
\ethm
\bprf
Proof in the full version of the paper at Appendix~\ref{sec:appendix}.
\eprf
\bnote \normalfont
We remark that interchanging the roles of initial and final codes results in a secure convertible code in the split regime for the parameters $\lambdaf \le \max(r^I, r^F)$.
\enote

 \paragraph*{AI use statement}
The authors used Claude Opus 5 and Gemini Edu to discuss and verify some of the proofs and results. Claude Opus 5 was used to generate an initial version of the Fig.~\ref{fig:convcode}. All mathematical statements, proofs, and conclusions were independently verified by the authors, who take full responsibility for the content of this paper. 

\bibliographystyle{IEEEtran}
\bibliography{secureconvertible}

\appendix \label{sec:appendix}
\blem\label{lem:nestedpropertygrs}
The $[n^I, k^I]$ initial code defined by generator matrix ${G}^I = \mathrm{V}_{k^I}(\mathcal{X} \cup \mathcal{Y})$ (see section~\ref{sec:grs}) is an $[n^I, k^I]$ $\ell$-nested code i.e., $P_{\kappa}\hat{G}^I$ is generator of an $[n^I, \ell]$-MDS code.
\elem
\bprf
From the definition of the initial code in the GRS based secure convertible code construction we have:
\bean
P_{\kappa}\hat{G}^I &=& \begin{bmatrix}
    \alpha_1^{k^I-\ell} & \alpha_2^{k^I-\ell} & \cdots & \alpha_n^{k^I-\ell}\\
    \vdots & \vdots & \ddots & \vdots\\
    \alpha_1^{k^I-1} & \alpha_2^{k^I-1} & \cdots & \alpha_n^{k^I-1}\\
\end{bmatrix}\\
&=& \text{V}_{\ell}(\alpha_1, \cdots, \alpha_n) \text{diag}(\alpha_1^{k^I-\ell}, \alpha_2^{k^I-\ell}, \cdots,  \alpha_n^{k^I-\ell})
\eean
Therefore clearly $P_{\kappa}\hat{G^I}$ is generator matrix of a $[k^I, \ell]$-MDS code.
\eprf

We assume that the systematic generator matrices of the initial and final codes are of the form $\hat{G}^I=[I_{k^I}\mid \hat{P}^I],
\quad
\hat{G}^F=[I_{k^F}\mid \hat{P}^F].$ Partition $\hat{P}^F$ into $\lambdai $consecutive row blocks,
each containing $k^I$ rows, and denote the i-th block by
$\hat{P}^{F,i}$.
By the parallel-block-reconstructability
property, each column of $\hat{P}^{F,i}$ is a nonzero scalar
multiple of a column of $\hat{P}^I$. This property is used to prove the $\ell$-secrecy property in Proof of Theorem~\ref{thm:GRSsecureconv}. 

Proof of Theorem~\ref{thm:GRSsecureconv}: We prove the $\ell$-secrecy property first by using Lemma~\ref{lem:Gjointsecrecycondition}. This along with the decoding property argued in Remark~\ref{rem:decod} imply the code is $(\ell, k_S=\lambda^I(k^I-\ell)$-secure. We do the proof for the case where $r^F \le r^I$. The proof for $r^I < r^F$, follows as the initial code considered in this case can be thought of a puncturing of $[\tilde{n}^I=k^I+r^F, k^I]$ code implying that the eavesdropper scenarios need to prove in this case are a subset of scenarios shown for $r^F \le r^I$ case. 


By Lemma~\ref{lem:Gjointsecrecycondition} it is enough to show that $P_{\kappa, D} \hat{G}_{\text{joint}}(:, \mathcal{B})$ has full rank given $\hat{G}_{\text{joint}}(:, \mathcal{B})$ has full rank where $B \subseteq [\lambdai n^I + \lambdaf n^F]$ is the index of symbols accessed by the eavesdropper. Let $\mathcal{B}_i \subseteq [n^I]$ be the index of symbols within $i$-th initial codeword accessed by the eavesdropper and $\mathcal{B}_0$ be the index of parity symbols accessed from the final codeword. We can assume that only parity symbols are accessed from the final codeword. Since eavesdropper can access only $\ell$ symbols we have that $\sum_{i=0}^{\lambdai} |\mathcal{B}_i| \le \ell$. Define:
\bean
\mathcal{B} = \cup_{i=1}^{\lambdai} \left( (i-1)  n^I+\mathcal{B}_i \right) \cup \left( \lambdai n^I + k^F+\mathcal{B}_0 \right).
\eean
Let $\un{x} \in \Null(P_{\kappa, D}\hat{G}_{\text{joint}}(:, \mathcal{B}))$ and  $\un{x}=(\un{x}_1^T~\un{x}_2^T~\cdots~\un{x}_{\lambdai}^T~\un{x}_0^T)$ where $\un{x}_i \in \mathbb{F}
_q^{|\mathcal{B}_i|}$. From the definition of $\hat{G}_{\text{joint}}(:, \mathcal{B})$ shown in equation~\eqref{eq:Gjoint} we get:
\bean
\scalebox{0.7}{$\left[ \begin{array}{cccc|cccc}
     P_{\kappa}\hat{G}^I(:, \mathcal{B}_1) & 0 & \cdots & 0 & P_{\kappa}\hat{P}^{F,1}(:, \mathcal{B}_0)  \\
     0 & P_{\kappa}\hat{G}^I(:, \mathcal{B}_2) & \cdots & 0 & P_{\kappa}\hat{P}^{F,2}(:, \mathcal{B}_0)  \\
     \vdots & \vdots & \ddots & 0 & \vdots   \\
     0 & 0 & \cdots &  P_{\kappa}\hat{G}^I(:, \mathcal{B}_{\lambdai}) & P_{\kappa}\hat{P}^{F,\lambdai}(:, \mathcal{B}_0)\\
\end{array}\right] \begin{bmatrix}
    \un{x}_1\\
    \un{x}_2\\
    \vdots\\
    \un{x}_{\lambdai}\\ 
    \un{x}_0
\end{bmatrix}=0$}.
\eean
It follows that for every $i \in [\lambdai]$:
\bean
\begin{bmatrix}
    P_{\kappa}\hat{G}^I(:,\mathcal{B}_i) & P_{\kappa}\hat{P}^{F,i}(:, \mathcal{B}_0)
\end{bmatrix} \begin{bmatrix}
    \un{x}_i\\
    \un{x}_0
\end{bmatrix}=0.
\eean
From the parallel block reconstructible property $\hat{P}^{F,i} = \hat{P}^{I}(:, \sigma_i([r^F])) D_i$ where $\sigma_i$ is a permutation of set $[r^I]$ and $D_i$ is a $r^F \times r^F$ diagonal matrix. Let $D_i = \mathrm{diag}(d_{i,1}, d_{i,2}, \ldots, d_{i,r^F})$ then
$D_i(\mathcal{B}_0) = \mathrm{diag}(d_{i,t})_{t \in \mathcal{B}_0}$. We therefore get
\bean
\begin{bmatrix}
    P_{\kappa}\hat{G}^I(:,\mathcal{B}_i) & P_{\kappa}\hat{P}^{I}(:, \sigma_i(\mathcal{B}_0)) D_i(\mathcal{B}_0)
\end{bmatrix} \begin{bmatrix}
    \un{x}_i\\
    \un{x}_0
\end{bmatrix}=0.
\eean
Let $\mathcal{P}_i=k^I + \sigma_i(\mathcal{B}_0))$ and 
$\mathcal{S}_i =  \mathcal{B}_i \cup \mathcal{P}_i$. We get: $
\hat{G}^I(:, \mathcal{P}_i) = \hat{P}^{I}(:, \sigma_i(\mathcal{B}_0))$.
Note that the union defining $\mathcal{S}_i$ need not be disjoint, since $\mathcal{B}_i$ may
contain parity indices lying in $\mathcal{P}_i$. Define
$E_i \in \mathbb{F}_q^{|\mathcal{S}_i| \times |\mathcal{B}_i|}$ and
$F_i \in \mathbb{F}_q^{|\mathcal{S}_i| \times |\mathcal{B}_0|}$, with rows indexed by
$m \in \mathcal{S}_i$, by
\bean
E_i(m, m') = \mathbf{1}[m = m'] \quad (m' \in \mathcal{B}_i),\\
F_i(m, t) = d_{i,t}\,\mathbf{1}[\,m = k^I + \sigma_i(t)\,] \quad (t \in \mathcal{B}_0),
\eean
so that $\hat{G}^I(:, \mathcal{S}_i) E_i = \hat{G}^I(:, \mathcal{B}_i)$ and
$\hat{G}^I(:, \mathcal{S}_i) F_i = \hat{P}^{I}(:, \sigma_i(\mathcal{B}_0)) D_i(\mathcal{B}_0)$.
Setting
\bean
\un{w}_i \; = \; E_i\, \un{x}_i + F_i\, \un{x}_0 \; \in \; \mathbb{F}_q^{|\mathcal{S}_i|},
\eean
the earlier equation simplifies to:
\bean
\begin{bmatrix}
    P_{\kappa}\hat{G}^I(:,\mathcal{S}_i)
\end{bmatrix} \begin{bmatrix}
    \un{w}_i
\end{bmatrix}=0.
\eean

By Lemma~\ref{lem:nestedpropertygrs}, $P_{\kappa}\hat{G}^I$ is a generator matrix of an
$[n^I, \ell]$-MDS code, so any $\ell$ of its columns are linearly independent; in
particular $P_{\kappa}\hat{G}^I(:, \mathcal{S}_i)$ has full column rank as $|\mathcal{S}_i| \le \ell$ and
\bean
\un{w}_i \; = \; 0 \qquad \text{for every } i \in [\lambdai].
\eean

It remains to conclude that $\un{x} = 0$. By the definitions of $E_i$ and $F_i$, the
$i$-th block of $\hat{G}_{\text{joint}}(:, \mathcal{B})\,\un{x}$ is
\bean
\hat{G}^I(:, \mathcal{B}_i)\, \un{x}_i + \hat{P}^{F,i}(:, \mathcal{B}_0)\, \un{x}_0
& =& \hat{G}^I(:, \mathcal{S}_i)\big( E_i \un{x}_i + F_i \un{x}_0 \big)\\
&=& \hat{G}^I(:, \mathcal{S}_i)\, \un{w}_i \; = \; 0 ,
\eean
and hence $\un{x} \in \Null\big(\hat{G}_{\text{joint}}(:, \mathcal{B})\big)$. Since
$\hat{G}_{\text{joint}}(:, \mathcal{B})$ is assumed to be of full rank, its null space is
trivial and $\un{x} = 0$. Therefore
$\Null\big(P_{\kappa, D}\hat{G}_{\text{joint}}(:, \mathcal{B})\big) = \{0\}$, i.e.\
$P_{\kappa, D}\hat{G}_{\text{joint}}(:, \mathcal{B})$ is of full rank, and by
Lemma~\ref{lem:Gjointsecrecycondition} the code is $\ell$-secure.

\eprf
\end{document}